 \documentclass[sigconf]{acmart}

\AtBeginDocument{%
  \providecommand\BibTeX{{%
    \normalfont B\kern-0.5em{\scshape i\kern-0.25em b}\kern-0.8em\TeX}}}

\copyrightyear{2024}
\acmYear{2024}

\usepackage{algorithm}
\usepackage{listings}
\usepackage{algorithmic}
\usepackage{xcolor}
\newcommand{\our}{{LogParser-LLM }}
\makeatletter
\newcommand*\ALG@numberline@{\refstepcounter{ALG@line}\nlset{\theALG@line}}
\makeatother
\begin{document}


\title{LogParser-LLM: Advancing Efficient Log Parsing with Large Language Models}


\author{Aoxiao Zhong}
\authornotemark[1]
\email{zhongaoxiao@gmail.com}
\orcid{0000-0002-7584-5476}

\affiliation{%
  \institution{Harvard University}
  \streetaddress{55 Fruit Street}
  \city{Cambridge}
  \state{MA}
  \country{US}
  \postcode{02114}
}
\affiliation{%
  \institution{Alibaba Group}
  \streetaddress{205 108th Ave NE, Suite 400}
  \city{Bellevue}
  \state{WA}
  \country{US}
  \postcode{98004}
}

\author{Dengyao Mo}

\email{dengyao.mo@alibaba-inc.com}

\affiliation{%
  \institution{Alibaba Group}
  \streetaddress{205 108th Ave NE, Suite 400}
  \city{Bellevue}
  \state{WA}
  \country{US}
  \postcode{98004}
}

\author{Guiyang Liu}

\email{wuming.lgy@alibaba-inc.com}

\affiliation{%
  \institution{Alibaba Group}
  \streetaddress{1008 Dengcai Street, Sandun Zhen, Xihu District}
  \city{Hangzhou}
  \state{Zhejiang}
  \country{China}
  \postcode{310030}
}

\author{Jinbu Liu}

\email{liujinbu.ljb@alibaba-inc.com}

\affiliation{%
  \institution{Alibaba Group}
  \streetaddress{1008 Dengcai Street, Sandun Zhen, Xihu District}
  \city{Hangzhou}
  \state{Zhejiang}
  \country{China}
  \postcode{310030}
}

\author{Qingda Lu}

\email{qingda.lu@alibaba-inc.com}

\affiliation{%
  \institution{Alibaba Group}
  \streetaddress{205 108th Ave NE, Suite 400}
  \city{Bellevue}
  \state{WA}
  \country{US}
  \postcode{98004}
}

\author{Qi Zhou}

\email{jackson.zhouq@alibaba-inc.com}

\affiliation{%
  \institution{Alibaba Group}
  \streetaddress{1008 Dengcai Street, Sandun Zhen, Xihu District}
  \city{Hangzhou}
  \state{Zhejiang}
  \country{China}
  \postcode{310030}
}

\author{Jiesheng Wu}

\email{jiesheng.wu@alibaba-inc.com}

\affiliation{%
  \institution{Alibaba Group}
  \streetaddress{205 108th Ave NE, Suite 400}
  \city{Bellevue}
  \state{WA}
  \country{US}
  \postcode{98004}
}

\author{Quanzheng Li}

\email{li.quanzheng@mgh.harvard.edu}

\affiliation{%
\department{CAMCA}
  \institution{Harvard Medical School,  Massachusetts General Hospital}
  \streetaddress{55 Fruit Street}
  \city{Boston}
  \state{MA}
  \country{US}
  \postcode{02114}
}

\author{Qingsong Wen}

\email{qingsongedu@gmail.com}

\affiliation{%
  \institution{Alibaba Group}
  \streetaddress{205 108th Ave NE, Suite 400}
  \city{Bellevue}
  \state{WA}
  \country{US}
  \postcode{98004}
}
\renewcommand{\shortauthors}{Aoxiao Zhong et al.}


\begin{abstract}

Logs are ubiquitous digital footprints, playing an indispensable role in system diagnostics, security analysis, and performance optimization. The extraction of actionable insights from logs is critically dependent on the log parsing process, which converts raw logs into structured formats for downstream analysis. Yet, the complexities of contemporary systems and the dynamic nature of logs pose significant challenges to existing automatic parsing techniques. The emergence of Large Language Models (LLM) offers new horizons. With their expansive knowledge and contextual prowess, LLMs have been transformative across diverse applications. Building on this, we introduce LogParser-LLM, a novel log parser integrated with LLM capabilities. This union seamlessly blends semantic insights with statistical nuances, obviating the need for hyper-parameter tuning and labeled training data, while ensuring rapid adaptability through online parsing. Further deepening our exploration, we address the intricate challenge of parsing granularity, proposing a new metric and integrating human interactions to allow users to calibrate granularity to their specific needs. Our method's efficacy is empirically demonstrated through evaluations on the Loghub-2k and the large-scale LogPub benchmark. In evaluations on the LogPub benchmark, involving an average of 3.6 million logs per dataset across 14 datasets, our LogParser-LLM requires only 272.5 LLM invocations on average, achieving a 90.6\% F1 score for grouping accuracy and an 81.1\% for parsing accuracy. These results demonstrate the method's high efficiency and accuracy, outperforming current state-of-the-art log parsers, including pattern-based, neural network-based, and existing LLM-enhanced approaches.

\end{abstract}





\maketitle
\renewcommand{\thefootnote}{}
\footnotetext{*This work was completed during an internship at Alibaba Cloud US.}
\section{Introduction}

Logs are pervasive records in the digital realm, vital for system diagnostics, security analysis, and performance optimization. As we navigate the complexities of contemporary digital environments, our systems, applications, and networks consistently generate vast amounts of logs. These abundant logs serve as an invaluable resource for understanding system behaviors, tracking activities, and uncovering hidden patterns. Their importance cannot be overstated, especially given the sophisticated nature of present-day systems and the crucial need for maintaining robust and efficient operations. This rich information source of cloud computing aids in tasks such as anomaly detection~\cite{zhang2019robust,du2017deeplog,xu2009detecting}, failure prediction~\cite{luo2021ntam,zhang2018prefix}, and failure diagnosis~\cite{zhang2021onion,he2018identifying}. In this digital age, effective utilization of logs can be the deciding factor between seamless operations and significant downtimes, highlighting their paramount importance in ensuring system reliability and security.

\begin{figure}[t]
  \centering
  \includegraphics[width=0.75\linewidth]{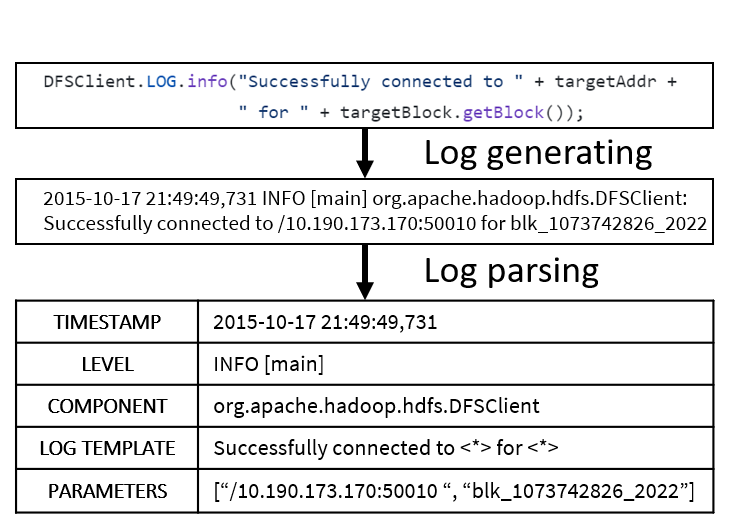}
  \caption{An example of log parsing.}
  
  \label{fig:example}
  
\end{figure}

Log parsing is a foundational step for many log-based diagnostic processes. Its main objective is to convert semi-structured log messages into a structured format, serving as the first step in a range of log analysis methods (e.g.,~\cite{lin2016log,chen2004failure,du2017deeplog}). This process entails identifying static components (referred to as log templates) and variable elements (known as log parameters) within log messages as shown in Figure ~\ref{fig:example}. Traditional methods for log parsing often involve comparing raw log messages to logging statements found in the source code~\cite{schipper2019tracing,pecchia2015industry,shang2012bridging} or creating regex patterns manually. However, with the growing volume and diversity of log messages, as well as the rapid evolution of modern software systems, these methods have become increasingly impractical~\cite{zhang2021onion}. Consequently, there has been a significant push towards developing automated log parsers, given their pivotal role in Artificial Intelligence for IT Operations (AIOps)~\cite{dang2019aiops,he2018identifying}.

To overcome the limitations of traditional log parsers, data-driven approaches have been developed by applying data mining techniques. Syntax-based methods were firstly developed, including frequent pattern mining~\cite{slct,lfa2010,logcluster2015,dai2020logram}, clustering~\cite{fu2009lke,tang2011logsig,hamooni2016logmine,mizutani2013shisho,shima2016lenma}, and heuristic-based~\cite{jiang2008ael,makanju2009iplom,he2017drain,du2016spell,yu2023brain,messaoudi2018molfi}. These methods operate on the principle that tokens remaining consistent across logs are likely templates, while those that differ are treated as parameters. The task becomes extracting the common parts from raw log messages. The lack of consideration for semantic meanings in logs leads to inaccurate identification of parameters, particularly for infrequently occurring logs. The reliance on hyper-parameters, such as pre-defined frequency thresholds or similarity thresholds, is a notable drawback of syntax-based methods. It necessitates careful tuning for specific log sources, thereby significantly restricting the parser's ability to generalize across diverse log data sources. To take the semantic meaning into consideration, recent studies utilize neural networks for log parsing. Uniparser~\cite{liu2022uniparser} uses LSTM aiming to build a universal parser that works for heterogeneous log data. LogPPT~\cite{le2023logppt} uses a pretrained transformer with few-shot learning. The necessity of labeled data for these methods, combined with their demonstrated underperformance on large-scale evaluations~\cite{jiang2023logpub}, renders them impractical to implement due to the scarcity of computing resources and labeled data.

Large Language Models (LLMs) have demonstrated remarkable capabilities across various domains. Given the vast pretraining datasets that encompass code and logging-related data, LLMs possess immense potential for log parsing. Pioneering research, as referenced in studies such as ~\cite{xu2023prompting, mudgal2023assessment, le2023evaluation, liu2023logprompt}, delved into LLM-based log parsing. However, the predominant focus of these studies has been on prompt engineering to enhance template extraction. Such methods parse logs line by line, incurring significant computational overhead due to the billions of parameters in LLMs. This renders these approaches somewhat impractical for broader applications.

To harness the capabilities of LLMs for practical log parsing, we introduce LogParser-LLM, which stands for \textbf{Log} \textbf{Parser} with \textbf{L}arge \textbf{L}anguage \textbf{M}odels. At its core, \our ~blends a prefix tree with an LLM-based template extractor. The latter capitalizes on the robustness of LLMs to semantically extract log templates from individual log messages. Concurrently, the prefix tree provides efficient log clustering grounded in syntax. On the one hand, the enhanced accuracy of the LLM template extractor ensures that the prefix tree is meticulously constructed and updated. On the other hand, the prefix tree aids in trimming the computational overhead of LLM by eliminating repetitive LLM calls. We have also integrated an automatic merging mechanism to rectify any template imperfections stemming from LLMs. These elements come together harmoniously to form a synergized parsing framework. Moreover, we exploit the in-context learning (ICL) capabilities of LLMs and implement named entity recognition (NER) prompting to further boost the accuracy of our LLM template extractor.

During the evaluation of our method, we encountered an intriguing observation. While our method generally produces satisfactory results, it doesn't always align with the annotated labels from the benchmark. This discrepancy can be attributed to what we term as the \textit{Granularity of Log Parsing}. Both the annotated labels and the model's outputs are logical in their own right. However, these differences in granularity can significantly impact existing metrics, as log entries parsed at differing granular levels are deemed incorrect. To more accurately quantify and understand this granularity discrepancy between parsing results, we introduce the metrics of \textit{Granularity Distance}. We further integrate human interactions in our method to allow users to calibrate the granularity based on their specific needs. 

We comprehensively evaluated our approach on the loghub-2k~\cite{zhu2019benchmark} and the extensive logPub~\cite{jiang2023logpub} datasets provided by the LogPAI team. Remarkably, \our surpasses previous state-of-the-art parsers, achieving a 48.3\% and 32.0\% increase in the F1 score for grouping and template accuracy, all without the need for domain-specific human effort. Notably, after calibrating granularity with ICL using a mere 32-shot labeled data for each domain, the performance enhancement reaches up to 56.8\% and 69.7\%. Given that each of the 14 datasets averaged 3.6 million logs, LLMs were only queried an average of 272.5 times, minimizing overhead and showcasing the feasibility of our method's real-world application.

The key contributions of this paper are summarized as follows:

\textbf{(1)} Introduction of \textit{\our}, a novel method leveraging LLMs for log parsing that merges syntactic and semantic insights, featuring an LLM template extractor and prefix tree to reduce LLM calls while processing millions of log lines efficiently.

\textbf{(2)} Enhancement of template extraction accuracy using ICL and NER prompting, characterized by low requirements for hyper-parameter tuning and labeled data, ensuring broad applicability and quick adaptation to new data.

\textbf{(3)} Development of new metrics for assessing parsing granularity, along with different options for users to adjust granularity effortlessly, making the parsing process more adaptable.

\textbf{(4)} Comprehensive validation of our approach through extensive testing on the \textit{loghub-2k} and \textit{logPub} benchmarks, demonstrating its effectiveness and efficiency, and confirming its suitability for addressing current challenges in log parsing.


\section{Related Work and Motivation}
Log parsing, extensively explored in research~\cite{zhu2019benchmark,khan2022guidelines}, identifies static \textbf{templates} and dynamic \textbf{parameters} within log entries. As shown in Figure~\ref{fig:example}, the template "Successfully connected to <*> for <*>" includes dynamic elements like "/10.190.173.170:50010" and “blk\_1073742826\_2022" as parameters. We categorize log parsing techniques into syntax-based, semantic-based, interactive, and LLM-based methods. We assess their pros and cons and identify opportunities for innovation, particularly in leveraging Large Language Models to improve log parsing capabilities.

\subsection{Syntax-based Log Parsers}
Syntax-based parsers detect templates by identifying repeating patterns as static and others as parameters. Frequency-based parsers like SLCT~\cite{slct}, LFA~\cite{lfa2010}, LogCluster~\cite{logcluster2015}, and Logram~\cite{dai2020logram}, build on token recurrence. Similarity-based parsers, including LKE~\cite{fu2009lke}, LogSig~\cite{tang2011logsig}, LogMine~\cite{hamooni2016logmine}, SHISO~\cite{mizutani2013shisho}, and LenMa~\cite{shima2016lenma} cluster logs by similarity. Heuristics-based parsers such as AEL~\cite{jiang2008ael}, IPLoM~\cite{makanju2009iplom}, Drain~\cite{he2017drain}, Spell~\cite{du2016spell}, Brain~\cite{yu2023brain}, and MoLFI~\cite{messaoudi2018molfi}, apply specific strategies including the longest common subsequence-based approach, iterative partitioning, prefix trees, and evolutionary algorithms for template extraction. These methods are fast and cost-efficient but may miss semantic details and require domain-specific tuning.

\subsection{Semantic-based Log Parsers}
Semantic parsers have evolved with neural networks like bidirectional LSTM, as seen in Semparser~\cite{huo2023semparser} and Uniparser~\cite{liu2022uniparser}, and pre-trained language models such as LogPPT~\cite{le2023logppt}. VALB~\cite{li2023valb} further enhances the model's semantic understanding by classifying specific parameter categories. These models require labeled data for training and classify tokens into templates or parameters. They offer semantic understanding and can generalize across log types, but also demand resource-intensive training and periodic updates, presenting significant operational challenges.

\subsection{Interactive Log Parsing}

Recent studies~\cite{wang2022spine,wang2023interactive} have incorporated user feedback into log parsers, facilitating human-in-the-loop log parsing. This approach not only enables the parser to swiftly adapt to evolving logs but also enhances the accuracy of template mining. 




\subsection{LLMs-based Log Parsing}
Large Language Models (LLMs) have emerged as transformative tools in numerous domains, demonstrating their prowess and versatility. Their pre-training on vast datasets, which include diverse content such as code and log data, makes them particularly adept for specialized tasks like log parsing. Studies like ~\cite{divlog2023, mudgal2023assessment, le2023evaluation, liu2023logprompt} have begun to tap into this potential, primarily focusing on prompt engineering to improve template extraction efficiency. While these advancements highlight the promise of LLMs in log parsing, they predominantly utilize a \textbf{line-by-line parsing approach}. This method, although innovative, leads to high computational demands due to LLMs' extensive parameter spaces, making these approaches \textbf{impractical for real-world applications} due to the significant computational overhead.

The benefits of LLMs extend beyond their raw computational ability, offering \textbf{deep semantic understanding} and the capacity to \textbf{generalize across different log formats}, adapting seamlessly to new data types. This adaptability is crucial, as it reduces the need for extensive preprocessing, hyper-parameter tuning, and manual labeling, streamlining the deployment process.

Despite these advantages, the practical deployment of LLMs in log parsing is hindered by their \textbf{high operational costs}. Effective utilization requires careful \textbf{prompt tuning}, a process that can be as resource-intensive as the computational demands of the models themselves. This challenge underscores the need for more efficient approaches that can leverage the strengths of LLMs without incurring prohibitive costs, ensuring their viability for broader, real-world application.

\section{Granularity of Log Parsing}

In this section, we delve into the granularity of log parsing. Starting with Section~\ref{Granu_character}, we characterize its two primary facets: Specificity and Applicability, elucidating them through an illustrative example. In Section~\ref{GD_measure}, we first highlight the shortcomings of existing metrics, emphasizing their inability to capture granularity nuances. Concluding the section, we introduce the granularity distance, a novel metric adept at gauging granularity discrepancies on two distinct levels, effectively addressing the gaps in prior metrics.

\subsection{Characterization of Granularity}\label{Granu_character}
The granularity of log parsing is pivotal for how the parsing result looks like. We primarily characterized the granularity by two dimensions: specificity and applicability.
\begin{figure}[t]
  \centering
  \includegraphics[width=\linewidth]{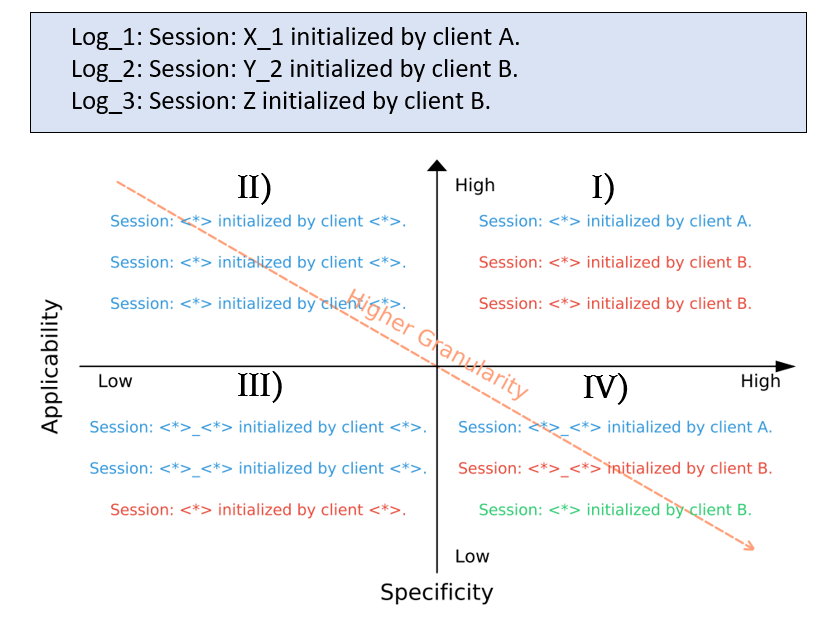}
  \caption{A demonstration of granularity variations in log parsing. Colors denote groups of templates. Applicability is represented on the vertical axis, while Specificity is represented on the horizontal axis.}

  \label{fig:granularity}
\end{figure}

\subsubsection{Specificity}

Specificity in log parsing indicates the depth of detail within a template. It is primarily driven by the \textbf{information and content} of templates. The more detailed they are, the higher the specificity.

\textbf{High Specificity (High Granularity):} Such templates have fewer, more detailed variable parts, aligning with a narrower set of logs due to their intricacy.

\textbf{Low Specificity (Low Granularity):} These are more general, with numerous variable components, catering to a broader log range.

The desired level of specificity often varies based on the log analysis context and user needs.

\subsubsection{Applicability}
Applicability in log parsing gauges a template's adaptability across varied log entries, primarily based on the \textbf{structure} of its placeholders. The more structurally generic they are, the broader their reach, translating to higher applicability.

\textbf{High Applicability (Low Granularity):} Templates here have a wide-reaching, generic structure, suitable for numerous logs.

\textbf{Low Applicability (High Granularity):} These are designed for specific log subsets, with unique structural placeholders.

Both specificity and applicability play crucial roles in determining the outcome of log templates, subsequently affecting metrics that measure grouping and parsing accuracy of log parsing. Together, they delineate the granularity of log parsing. The ideal granularity often finds a midpoint between these two dimensions and is shaped by user preferences and the nuances of individual use cases. Notably, even a minor discrepancy in granularity can result in substantially different groupings, a difference that can be exaggerated when using inappropriate metrics. This highlights the pressing need for a well-conceived metric. It is essential to recognize the inherently subjective nature of granularity. As such, it is inappropriate to strictly label a particular granularity as dominant or to view benchmark dataset labels as definitive standards. Figure~\ref{fig:granularity} illustrates the parsing outcomes at varying granularities, using three representative log messages from the Windows dataset in loghub-2k. Applicability is shown through session names, while specificity is shown through client names. High applicability and low specificity result in a generic structure that matches more log entries, indicating lower granularity. The benchmark uses granularity I) as its labeled ground truth.

\subsection{Measuring Granularity Discrepancy}\label{GD_measure}

Existing evaluation metrics, while versatile, emphasize either the accuracy of grouping logs or the fidelity in extracting templates and parameters. Both dimensions are indispensable, especially considering their implications for downstream tasks like log anomaly detection. However, these metrics often overlook the subtle granularity differences inherent in log parsing. Existing benchmark datasets~\cite{zhu2019benchmark,jiang2023logpub} are anchored to the annotators' subjective interpretations, suggesting that multiple valid granular interpretations can exist for a single log. Such diversity challenges the conventional wisdom of treating annotated labels as an unequivocal gold standard. Instead of a myopic focus on exact matches, a more encompassing metric that can quantify and understand this granularity discrepancy is imperative. 

\subsubsection{Existing metrics} 
We examine four prevalent metrics in this section. The widely recognized message-level metrics, Grouping Accuracy (GA)~\cite{zhu2019benchmark} and Parsing Accuracy (PA)~\cite{dai2020logram}, focus on the volume of messages associated with each template, often prioritizing templates with a larger number of log messages. To address this bias, template-level metrics like F1-score of Group Accuracy (FGA)~\cite{jiang2023logpub} and F1-score of Template Accuracy (FTA)~\cite{khan2022guidelines} have been introduced, ensuring an equitable evaluation of each template. The detailed definitions can be found in Appendix~\ref{app_metrics}.

GA and PA primarily evaluate based on the volume of log messages, making them susceptible to biases from imbalanced templates. In real-world scenarios, less frequent templates, such as error messages, might be of paramount importance. Their misinterpretation could be detrimental, yet this might not be reflected effectively using these metrics. Template-level metrics ensures a holistic evaluation of log parsers, giving equal importance to each template. However, while these metrics minimize biases from frequent templates, they still present challenges. If a token is interpreted differently based on granularity nuances, whether designated as a static part or a parameter, it might result in considerable variances in template counts. Additionally, such metrics don't provide a clear insight into granularity differences.

\subsubsection{Granularity Distance (GD)}
\label{sec:gd}
In light of the discussions above and the sensitivity of existing metrics to subtle granularity discrepancies, we introduce the \textit{Granularity Distance} metric. Inspired by the traditional edit distance, this metric calculates \textit{the minimum operations necessary to transform one parsing result into another}. It serves as a quantitative reflection of the least human intervention needed to attain the desired granularity. This metric can be dissected into two main components:

\textbf{Grouping Granularity Distance(GGD)}: 
This aspect emphasizes the grouping of log messages. The aim is to match the expected grouping of log messages without mandating identical templates within those groups.

\textbf{Parsing Granularity Distance(PGD)}: 
This is a more rigorous metric requiring an exact match for each log template. Disparities in the parsed templates increment the distance.

For the operations contributing to this distance:

\textbf{Operations on GGD}:
\textit{1) Merge}: Combine groups by changing one static section to variable.
\textit{2) Split}: Separate groups by switching one variable to static section.

\textbf{Operations on PGD}:
 \textit{1) Static to Variable}: Convert a static section of the template to a variable.
 \textit{2) Variable to Static}: Revert a variable within the template to a static section.
 
Similar to the edit distance, granularity distance possesses symmetrical properties, meaning the distance from one log template to another is the same as the distance from the second to the first. Additionally, granularity distance satisfies the properties of non-negativity and identity of indiscernibles, akin to the traditional metrics in distance measurement. This ensures a consistent and logical comparison of log parsing granularity between different parsing results.

It is straightforward to compute GD when logs are accurately tokenized, and each token is categorized as either a parameter or a template. However, such precise labeling and tokenization are often absent. To circumvent this, an approximate version of GGD can be derived by merely tallying the merge and split operations required to transition from one grouping to another.

\section{Methodology}
In this section, we introduce \textbf{LogParser-LLM} tailored to tackle the challenges previously highlighted. Our approach is built upon four key pillars:
\textbf{1) Enhanced Template Extraction}: Leveraging the prowess of LLMs, we aim to boost the accuracy of template extraction.
\textbf{2) Efficient LLM Use}: We design an algorithm that harnesses the advanced capabilities of LLMs while optimizing resource consumption.
\textbf{3) Reduced Human Effort with Broad Applicability}: Our method minimizes human intervention, especially in label annotation and hyper-parameter tuning, yet remains versatile across various domains and log formats.
\textbf{4) Interactive Feedback Integration}: Our method is integrated with human feedback for parsing granularity calibration.
The following sections delve deeper into these principles, elucidating the techniques and decisions underpinning our approach.

\subsection{Preprocessing}

Our method hinges on minimal preprocessing, using only a basic regular expression to extract log content. While many approaches demand greater domain knowledge, often employing regular expressions to substitute common variables like IP addresses and block IDs~\cite{he2017drain}, we retain the original message, ensuring the LLM grasps the log's full context. Unlike other strategies that use distinct separators for log tokenization~\cite{yu2023brain,liu2022uniparser,fu2022investigating}, we consistently tokenize using spaces. Hence, unless otherwise specified, tokens in following sections are space-separated, capitalizing on the LLM's native tokenizer. This streamlined preprocessing minimizes the need for specialized expertise, yet upholds strong log parsing efficacy.

\subsection{Base Algorithm with Prefix Parse Tree}
Central to our methodology is a base algorithm employing a prefix parse tree, inspired by the efficiency demonstrated in Drain~\cite{he2017drain}. This section elaborates on the data structures integral to the algorithm, detailing their design and their roles in addressing the aforementioned principles. Specifically, we'll elucidate how incoming logs are matched with existing clusters during tree traversal, how and when LLMs are invoked for template extraction, and the dynamics of updating the tree with new templates obtained from the LLM extractor.

\subsubsection{Data Structures}
\begin{figure*}[t]
  \centering
  \includegraphics[width=0.8\linewidth]{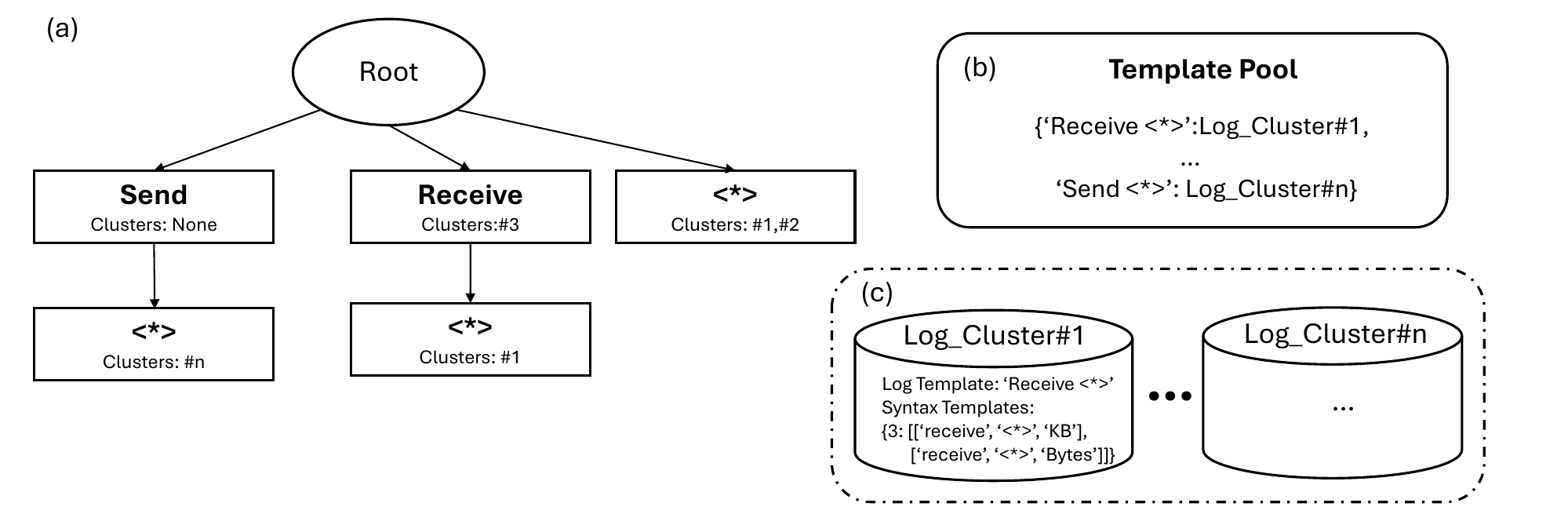}
\caption{An example demonstrating the data structures in our method: (a) A prefix parse tree with nodes linking to log clusters, (b) a Template Pool mapping log templates to log clusters, and (c) Log Clusters containing collections of logs with the same log template.}
  \label{fig:structure}
\end{figure*}
Three primary data structures form the backbone of our approach: a set of log clusters, a template pool, and a prefix parse tree. Figure~\ref{fig:structure} offers a visual representation of this organizational structure. The subsequent discussion delineates their respective functionalities:  

\textbf{Log Cluster:}
A log cluster is a collection of logs with the same template. It keeps track of the individual log IDs and stores a log embedding, created by an LLM encoder, for future use. Each cluster is characterized by its \textit{log template}, extracted via LLM, and possibly multiple \textit{syntax templates} aiding the prefix tree in its traversal and template matching processes. While \textit{syntax templates} correspond directly with the tokens of the raw logs, identifying static and variable parts, the \textit{log templates} from the LLM may represent several tokens with a single placeholder. These \textit{syntax templates} are stored in a dictionary, utilizing token counts as keys and corresponding template lists as values.

\textbf{Template Pool:}
The \textit{template pool} establishes a linkage, mapping \textit{log templates} to their respective \textit{log clusters}.

\textbf{Prefix Parse Tree:}
In this tree structure, every node—bar the root—symbolizes a token. The wildcard token "<*>" serves as a universal matcher for any token. Crucially, not just the leaf nodes, but virtually all nodes (excluding the root) can possess pointers to log clusters matching the token sequence extending from the root. A unique feature to note is that a single log cluster might be accessible from multiple nodes, courtesy of the potential existence of various syntax template variants for a given log cluster.

\subsubsection{cluster matching with tree search}
Upon receiving a new log, our first step is tokenization. Tokens are then processed sequentially, with each token checked against nodes in the prefix tree. After matching the initial token, we proceed to the subsequent token, considering only the children of the previously matched node. This progression continues either until all tokens are matched or when no further matching tokens exist. Throughout this traversal, log clusters referenced by the encountered nodes are shortlisted as potential candidates for a thorough match evaluation. 

At this juncture, we have pinpointed a subset of log clusters consistent with the rules encoded in the tree path. Our task now is to determine the genuine match from these candidates. Contrary to existing methodologies such as those in~\cite{he2017drain,yu2023brain}, which deploy similarity metrics and predefined, dataset-specific thresholds, our approach crystallizes outcomes into three distinct categories: i) Strict match, ii) Loose match, and iii) No match. For each prospective cluster, an initial check compares the token count between the incoming log and the cluster's syntax templates. Discrepant token counts immediately exclude the possibility of a match. Following this, a 'loose match' is attempted, aligning tokens from the syntax template and the log. Here, any token within the syntax template containing the "<*>" wildcard can align with any log token. To illustrate, a token such as "prefetching...<*>" can loosely align with any log entry with a singular token. After achieving a loose match, regular expressions ensure a rigorous alignment with elements outside the "<*>" in the syntax template. A complete token alignment signifies a strict match. It is worth noting that the matching process stops upon achieving a strict match. In scenarios where a strict match is identified, the log is straightforwardly added to the matched cluster. Conversely, in the absence of a strict match, the LLM template extractor is invoked for template extraction, followed by the necessary updates to the data structures.

Our method's precision, rooted in the capabilities of LLMs, eliminates the need for meticulous hyperparameter tuning across log sources. Leveraging LLMs' proficiency, which typically yields semantically accurate templates for individual logs, facilitates this stringent matching paradigm. Not only does it simplify the tuning process, but it also optimizes the number of calls to the LLM. Ideally, if we operate under the assumption that LLMs generate templates mirroring the ground truth, the volume of LLM calls is effectively capped at the total number of distinct syntax templates. This count is in the ballpark of the total number of log templates, offering a scalable approach.

\subsubsection{Parse tree update}
The comprehensive update rule is elucidated in Algorithm~\ref{alg:our}.  As outlined in lines 8-9, for scenarios of either a loose match or no match, the LLM is invoked to derive a log template. If this extracted template already resides in the template pool, it suggests that the current log pertains to an existing cluster but with an alternative syntax template variant. In such cases, the associated cluster can be swiftly identified via the template pool mapping. It then becomes essential to integrate the novel syntax template into the cluster and adjust the tree to accommodate nodes that align with this new syntax template.

Conversely, if the template isn't found in the template pool yet a loose match has been identified, the LLM is once more consulted. Its task here is to determine if the loosely matched cluster can integrate this new log. A positive outcome leads to the generation of a merged template. Subsequently, both the syntax and log templates of the cluster undergo an update, with the merged template being added to the template pool.

If a log, after undergoing the entire aforementioned process, still has not been allocated to an existing cluster, it is indicative of a unique log template. Such instances mandate the creation of a new cluster, with corresponding updates made to the tree.

\begin{algorithm}[tb]
   \caption{\textsc{\our} }
   \label{alg:our}
\begin{algorithmic}[1]
    \item[] {\bfseries Input:} $logs$, $root\_node \quad$  {\bfseries Result:} $log\_clusters$, $tree$
    \STATE $log\_clusters \leftarrow \{\}$
    \STATE $template\_pool \leftarrow \{\}$
    \FOR{$log \textbf{ in } logs$}
        \STATE $matched\_clusters \gets \text{search}(tree,log)$
        \IF{strict\_match}
            \STATE $strict\_matched\_cluster.\text{add}(log)$
            \STATE $added \gets \text{True}$
        \ELSIF{loose\_match \textbf{or} no\_match}
            \STATE $template \gets \text{get\_llm\_template}(log,log\_clusters)$
            \IF{$template$ \textbf{in} $template\_pool$}
                \STATE $\text{update\_tree}(tree,log,template\_pool[template])$
                \STATE $added \gets \text{True}$
            \ELSE             
                \FOR{$cluster$ in $loose\_matched\_clusters$}
                    \STATE check\_merge($log, cluster$)
                    \IF{merge}
                        \STATE $cluster.\text{update}(merged\_template)$
                        \STATE $cluster.\text{add}(log)$
                        \STATE$template\_pool[merged\_template]=cluster$
                        \STATE \textbf{break}
                        \STATE $added \gets \text{True}$
                    \ENDIF
                \ENDFOR

            \ENDIF
        \ENDIF
        \IF{\textbf{not} $added$}
            \STATE $new\_cluster\gets  \text{create\_cluster}(log,template)$
            \STATE $\text{update\_tree}(tree,log,new\_cluster)$
            \STATE $template\_pool[template]=cluster$
        \ENDIF
    \ENDFOR
\end{algorithmic}

\end{algorithm}

\subsection{Enhancing LLM Template Extraction}
While the base algorithm already paves the way for efficient and precise log cluster matching, there remains room to refine the accuracy of the LLM template extractor. To this end, we introduce variable-aware prompting, amalgamating it with in-context learning. This fusion not only amplifies the LLM's task comprehension but also augments its overall performance.

Additionally, there exist other straightforward avenues to bolster template extraction capabilities. One could leverage more powerful LLMs available in the ever-evolving landscape of language models. Alternatively, supervised fine-tuning of an LLM using labeled data presents another viable strategy.
\subsubsection{Variable-Aware Prompting}

Past research~\cite{li2023valb} has highlighted the benefits of identifying and classifying specific variables within logs. By categorizing these variables, not only is the accuracy of template extraction enhanced, but it also proves advantageous for subsequent tasks. Drawing inspiration from this research and the concept of chain-of-thought prompting~\cite{wei2022chain}, we restructure our prompts. These prompts now serve dual purposes: they identify variables and categorize them into one of the ten classifications as outlined in~\cite{li2023valb}. This refined approach prompts the model to understand and determine which components should be classified as variables and the reasoning behind such categorization.

\subsubsection{In-Context Learning with K-Shot Demonstrations}

In-context learning (ICL) has become a favored approach when using LLMs for downstream tasks without the need for finetuning~\cite{dong2022icl}. Typically, ICL-based prompts contain three elements:
\textit{Instruction}: A task-specific description.
\textit{Demonstrations}: A set of examples, essentially pairs of queries coupled with their ground truth answers.
\textit{Query}: The direct question to which the LLM provides a response.
Each time the LLM is called upon for template extraction, we draw a sample of $k=3$ examples from our existing pool of log template pairs. Incorporating the principles of Variable-Aware Prompting, we include ten examples, each representing a distinct type of log parameter, as seed examples. Subsequent template extraction results expand this pool. To obtain these samples, we calculate cosine similarity between LLM embedding of query log and all embeddings present in the example pool. The top-k samples are then chosen as k-shot demonstrations within the prompt.

\subsection{Optimal Granularity via Human-in-Loop}

Integrating human expertise into the automated log parsing process is key to achieving the right granularity. Human input can be seamlessly incorporated at various stages of the parsing pipeline to enhance accuracy and maintain consistency:

\textbf{1) Pre-Processing Intervention:} Experts annotate a sample of logs before parsing begins. These annotations serve dual purposes: they can be used as seed examples for In-Context Learning (ICL) or to fine-tune LLMs, ensuring the model's output aligns more closely with specific parsing needs.

\textbf{2) Real-Time Calibration:} During the parsing process, human judgment can be applied to guide decisions on template merging, ensuring the parsing maintains the desired level of granularity throughout.

\textbf{3) Post-Processing Refinement:} After parsing, the system identifies potential merges or splits based on semantic similarity or template variability. Experts review these suggestions, making adjustments to achieve the optimal granularity.

In Table~\ref{tab:logpub}, we demonstrate how LogParser-LLM-C incorporates pre-processing intervention, enhancing the base LogParser-LLM's capabilities. For real-time calibration, human expertise can be used to refine the merging process in line with desired granularity levels, as outlined in line 15 of Algorithm 1. Post-processing refinement can integrate methods like those suggested in \cite{wang2023interactive} for effective final adjustments, ensuring parsed logs accurately reflect the intended granularity.





\section{Experiments}
We assess the effectiveness of our method using two datasets: loghub-2k~\cite{zhu2019benchmark} and logPub~\cite{jiang2023logpub}. First, we detail the experimental settings. Subsequently, we outline the evaluation metrics employed, highlighting a novel metric we introduce to gauge the granularity distance of parsing outcomes. In examining results from the loghub-2k dataset, our primary objective is to elucidate the contribution of each design component of our method. With the logPub benchmark, our intent is to demonstrate both the effectiveness and efficiency of our approach when handling large-scale datasets in practice.

\subsection{Experimental Settings}
\begin{table*}[t]
\footnotesize
\setlength{\tabcolsep}{1pt}
  \caption{Comparison of various log parser algorithms on large-scale logPub dataset. The best results are in bold.}
  
  \label{tab:logpub}
  \begin{tabular}{c|cccccc|cccccc|cccccc|cccccc|cccccc}
  \toprule
 & \multicolumn{6}{c|}{Drain} & \multicolumn{6}{c|}{Uniparser}& \multicolumn{6}{c|}{LogPPT}&\multicolumn{6}{c|}{LogParser-LLM}&\multicolumn{6}{c}{LogParser-LLM-C}\\
 & GA& PA& FGA& FTA &GGD&PGD
& GA& PA& FGA& FTA &GGD&PGD
& GA& PA& FGA& FTA &GGD&PGD
& GA& PA& FGA& FTA & GGD&PGD
& GA& PA& FGA& FTA& GGD&PGD\\
 \midrule

Proxifier   & 69.2           & 68.8 & 20.6  & 17.6 & 4          & 14    & 50.9           & 63.4          & 28.6 & 45.7 & 5    & 10    & \textbf{98.9} & \textbf{100.0} & \textbf{87.0} & \textbf{95.7} & \textbf{1} & \textbf{1} & 51.0           & 63.4 & 40.0           & 53.3 & 5          & 9     & \textbf{98.9}  & \textbf{100.0} & \textbf{87.0}  & \textbf{95.7} & \textbf{1}    & \textbf{1}    \\
Linux       & \textbf{68.6}  & 11.1 & 77.8  & 25.9 & 30         & 432   & 28.5           & 16.4          & 45.1 & 23.2 & 108  & 274   & 20.5          & 16.8           & 71.2          & 42.8          & 29         & 104        & 27.0           & 16.3 & 80.1           & 46.6 & 18         & 81    & 53.4           & 49.4           & \textbf{91.1}  & \textbf{74.0} & \textbf{10}   & \textbf{68}   \\
Apache      & \textbf{100.0} & 72.7 & 100.0 & 51.7 & \textbf{0} & 21    & 94.8           & 94.2          & 68.7 & 26.9 & 11   & 31    & 78.6          & 94.8           & 60.5          & 36.8          & 6          & 23         & \textbf{100.0} & 85.7 & \textbf{100.0} & 65.5 & \textbf{0} & 8     & \textbf{100.0} & \textbf{99.5}  & \textbf{100.0} & \textbf{82.8} & \textbf{0}    & \textbf{5}    \\
Zookeeper   & 99.4           & 84.3 & 90.4  & 61.4 & 2          & 30    & 98.8           & \textbf{98.8} & 66.1 & 51.0 & 14   & 31    & 96.7          & 84.5           & 91.8          & 80.9          & 4          & 10         & 98.8           & 81.9 & 86.2           & 72.4 & 2          & 19    & \textbf{99.5}  & 96.8           & \textbf{92.9}  & \textbf{85.7} & \textbf{1}    & \textbf{13}   \\
Hadoop      & 92.1           & 54.1 & 78.5  & 38.4 & 18         & 210   & 69.1           & 88.9          & 62.8 & 47.6 & 38   & 119   & 48.3          & 66.6           & 52.6          & 43.4          & 46         & 81         & 93.8           & 67.6 & 87.3           & 55.0 & 14         & 108   & \textbf{94.5}  & \textbf{90.6}  & \textbf{88.9}  & \textbf{81.0} & \textbf{11}   & \textbf{41}   \\
HealthApp   & 86.2           & 31.2 & 1.0   & 0.4  & 11         & 138   & 46.1           & 81.7          & 74.5 & 46.2 & 16   & 60    & 99.8          & \textbf{99.7}  & 94.7          & 82.2          & 4          & 8          & 99.8           & 58.2 & 95.6           & 81.8 & 4          & 5     & \textbf{100.0} & 98.2           & \textbf{96.5}  & \textbf{89.0} & \textbf{3}    & \textbf{7}    \\
OpenStack   & 75.2           & 2.9  & 0.7   & 0.2  & 6618       & 23**   & \textbf{100.0} & 51.6          & 96.9 & 28.9 & 1    & 7     & 53.4          & 40.6           & 87.4          & 73.8          & 4          & 4          & \textbf{100.0} & 49.6 & \textbf{100.0} & 79.2 & \textbf{0} & 11    & \textbf{100.0} & \textbf{100.0} & \textbf{100.0} & \textbf{97.9} & \textbf{0}    & \textbf{1}    \\
HPC         & 79.3           & 72.1 & 30.9  & 15.2 & 10         & 178   & 77.7           & 94.1          & 66.0 & 35.1 & 10   & 58    & 78.2          & 99.7           & 78.0          & 76.8          & 12         & 31         & \textbf{86.4}  & 94.2 & 76.0           & 72.6 & \textbf{6} & 180   & \textbf{86.4}  & \textbf{99.8}  & \textbf{76.8}  & \textbf{74.6} & \textbf{6}    & \textbf{29}   \\
Mac         & 76.1           & 35.7 & 22.9  & 6.9  & 102        & 1347  & 73.7           & 68.8          & 69.9 & 28.3 & 73   & 624   & 54.4          & 39.0           & 49.3          & 27.4          & 177        & 489        & 89.7           & 30.3 & 84.7           & 36.2 & 42         & 444   & \textbf{91.5}  & \textbf{76.4}  & \textbf{86.4}  & \textbf{60.6} & \textbf{33}   & \textbf{297}  \\
OpenSSH     & 70.7           & 58.6 & 87.2  & 48.7 & 3          & 33    & 27.5           & 28.9          & 0.9  & 0.5  & 15   & 26    & 27.7          & 65.4           & 8.1           & 10.5          & 17         & 26         & \textbf{78.0}  & 69.0 & \textbf{96.1}  & 88.3 & \textbf{1}          & 9     & \textbf{78.0}  & \textbf{100.0} & \textbf{96.1}  & \textbf{98.7} & \textbf{1}    & \textbf{2}    \\
Spark       & 88.8           & 39.4 & 86.1  & 41.2 & 18         & 239   & 85.4           & 79.5          & 2.0  & 1.2  & 62   & 186   & 47.6          & 95.2           & 37.4          & 29.9          & 75         & 221        & \textbf{97.6}  & 80.2 & 85.2           & 46.3 & 16         & 148   & \textbf{97.6}  & \textbf{99.7}  & \textbf{88.2}  & \textbf{68.1} & \textbf{11}   & \textbf{101}  \\
Thunderbird & 83.1           & 21.6 & 23.7  & 7.1  & 137        & 2043  & 57.9           & 65.4          & 68.2 & 29.0 & 194  & 976   & 56.4          & 40.1           & 21.6          & 11.7          & 282        & 1012       & 73             & 57.1 & 80.0           & 56.0 & 104        & 662   & \textbf{67.5}  & \textbf{64.3}  & \textbf{83.1}  & \textbf{59.3} & \textbf{88}   & \textbf{615}  \\
BGL         & 91.9           & 40.7 & 62.4  & 19.3 & 48         & 434   & 91.8           & 94.9          & 62.4 & 21.9 & 43   & 209   & 24.5          & 93.8           & 25.3          & 26.1          & 69         & 164        & \textbf{93.8}  & 81.0 & 78.9           & 50.0 & 34         & 154   & 88.9           & \textbf{97.6}  & \textbf{84.0}  & \textbf{71.6} & \textbf{24}   & \textbf{85}   \\
HDFS        & 99.9           & 62.1 & 93.5  & 60.9 & 2          & 6     & \textbf{100.0} & 94.8          & 96.8 & 58.1 & 1    & 1     & 72.1          & 94.3           & 39.1          & 31.2          & 18         & 59         & \textbf{100.0} & 94.8 & 74.7           & 57.8 & 5          & 26    & \textbf{100.0} & \textbf{100.0} & \textbf{96.8}  & \textbf{96.8} & \textbf{1}    & \textbf{1}    \\
\midrule
Average     & 84.3           & 46.8 & 55.4  & 28.2 & 500.2      & 394.2 & 71.6           & 73.0          & 57.8 & 31.7 & 42.2 & 186.6 & 61.2          & 73.6           & 57.4          & 47.8          & 53.1       & 159.5      & \textbf{90.9}  & 68.3 & 85.7           & 63.1 & 17.9       & 133.1 & 89.7           & \textbf{90.9}  & \textbf{90.6}  & \textbf{81.1} & \textbf{13.6} & \textbf{90.4}\\

\bottomrule
\end{tabular}
\end{table*}

\begin{table*}[t]
\footnotesize
    \centering
        \caption{Efficiency and effectiveness of the LogParser-LLM with different LLMs. The best results are in bold.}
    \begin{tabular}{l|c|ccc|cccccc}
    \toprule
 & Avg. \# of & \multicolumn{3}{|c}{Avg. Time(s)}& \multicolumn{5}{|c}{Avg. Metrics}\\
         &  LLM Calls&  Per Infer.&  Base&  Total&  GA&  PA&  FGA&  FTA& GGD&PGD\\
    \midrule
         LogParser-LLM w/ GPT-3.5-turbo& 566.4 &0.52  & 522.88  &817.33& 82.0 &64.5  & 81.6 &  58.3 & 26.9&198.9\\
         LogParser-LLM-C w/ GPT-3.5-turbo (32shot)& 289.7 &0.52  & 461.67 & \textbf{612.31} &  \textbf{91.3}&  90.5&  90.2&  77.5& 15.4&120.7\\
         LogParser-LLM with GPT-4& 427.2 &4.18  & 452.15 & 2237.85 &  90.9&  68.3&  85.7&  63.1& 17.6&133.1
\\
         LogParser-LLM-C w/ GPT-4 (32shot) &\textbf{272.5}  & 4.18 & 433.22 &1572.27  &  89.7&  \textbf{90.9}&  \textbf{90.6}&  \textbf{81.1}& \textbf{13.6}&\textbf{90.4}\\
         LogParser-LLM w/ fine-tuned Llama-2-13b(32shot)&  6620.3& 2.54  & 621.33 &  2272.63&  78.5&  72.0&  66.8&  49.9& 45.9&194.0\\
    \bottomrule
    \end{tabular} 

    \label{tab:llms}
    
\end{table*}

\subsubsection{Datasets}
Loghub-2k is a widely recognized benchmark in the field of log parsing. It encompasses logs from 16 diverse systems, including distributed systems, supercomputers, operating systems, mobile platforms, server applications, and individual software packages. For every system source, 2,000 log messages are meticulously annotated. Complementing this, LogPub is a more recent, expansive iteration of Loghub-2k. It features 14 systems, with each averaging a substantial 3.6 million log lines, and showcases a pronounced increase in the number of log templates. This dataset offers a realistic, large-scale environment, paving the way for comprehensive evaluations of log parsing methodologies.

\subsubsection{Implementation Details} 
Our experimental setup involves a server powered by Ubuntu 20.04.3 LTS with 512GB of RAM. We use both ChatGPT (version gpt-3.5-turbo-0301) and GPT-4 (version gpt-4-0613) for template extraction. For embedding the logs, the text-embedding-ada-002 method is adopted. All interactions with these models are facilitated through the official OpenAI API. To guarantee consistency in our findings and support reproducibility, we maintain the temperature parameter at 0 to minimize variability. For fine-tuning our LLM, the Llama-2-13b model~\cite{touvron2023llama} serves as the foundation. Comprehensive details regarding this fine-tuning process can be found in Appendix B.2.
For in-context learning, we uniformly sample 32 log-template pairs from the first 10\% of each dataset based on token length as candidate logs. The same samples are employed for fine-tuning.

\subsection{Evaluation Metrics}
In alignment with prevailing methods outlined in~\cite{liu2022uniparser,khan2022guidelines,jiang2023logpub}, we utilize the GA, PA, FGA, PTA, RTA and FTA metrics delineated in Appendix~\ref{app_metrics} for evaluation. Furthermore, we use the Grouping Granularity Distance (GGD) proposed in Section~\ref{sec:gd} as a more intuitive metric to gauge the granularity discrepancies in parsing outcomes.

\subsection{Evaluation on Loghub-2k}
Our primary objective in conducting experiments with the smaller-scale Loghub-2k dataset is to assess the efficacy of our method's key components. Additionally, we employ this dataset as a development set, refining our prompts for LLM template extraction, merging check and verification. The final prompts we adopted are presented in Appendix. B. The most effective configurations determined through Loghub-2k are subsequently applied unchanged to the LogPub dataset for evaluation. 
\begin{table}[h]
  \caption{Comparison with existing LLM-based method on Loghub-2k.}
  \label{tab:comp}
  \footnotesize
  \begin{tabular}{llcccc}
  
    \toprule
    &\# of labeled logs&	GA&	PA	&PTA&RTA\\
    \midrule
 Eval of chatgpt\cite{le2023evaluation}& 0& 72.1 &54.3 &/&/\\
 & 4& 76.1 &79.0 &/&/\\
 \midrule 
 DivLog\cite{divlog2023}& 200& 92.8 &98.1&92.0&92.9\\
    \midrule
    LogParser-LLM& 0  &91.8	& 69.9  &	67.8&	66.8\\
                 & 4  &92.1	&74.7	&72.6	&74.9\\
                 & 32 &94.8	&90.5	&84.9	&85.3\\
                 & 200&95.9	&98.0	&96.8	&96.9\\
  \bottomrule
\end{tabular}
\end{table}

\textbf{Comparison with Existing LLM-based Parsers}
Existing LLM-based parsers, which process logs line-by-line, are impractical for evaluation on the expansive LogPub dataset due to the immense number of LLM calls required. We therefore use the Loghub-2k dataset for comparison, but advise caution in interpreting these results because of the dataset's limited scope and the possibility that a few well-chosen labeled samples might cover the majority of templates. In Table~\ref{tab:comp}'s results, our method either matches or exceeds the performance of existing approaches with an equivalent number of labeled logs, highlighting our method's effective use of LLMs for log parsing despite the constraints.

\textbf{Accuracy Evaluation}
We compare our model to three state-of-the-art methods: two syntax-based methods, Drain~\cite{he2017drain} and Brain~\cite{yu2023brain}, and one semantic-based method, LogPPT~\cite{le2023logppt}. As indicated in Table ~\ref{tab:ablation}, the previous state-of-the-art methods achieved higher GA and PA values because they were meticulously tuned with hyperparameters on each dataset to optimize these metrics. However, these values alone do not necessarily indicate superior performance. When evaluating with template-level metrics such as FGA and FTA, as well as our proposed GGD, our model outperforms them without the need for any domain-specific configuration.

\textbf{Ablation} The results of our ablation study for different components, including in-context learning (ICL), variable-aware prompt (VA), and automatic template merge (Merge), are presented in Table~\ref{tab:ablation}. The numbers clearly demonstrate that each proposed component positively impacts the method's performance, as evidenced by the reduction in GGD. Notably, the most significant performance boost comes from the transition from GPT-3.5 to GPT-4. Furthermore, the enhancements from other components are even more pronounced with GPT-4, underscoring the potency of more powerful LLMs. Using GPT-4 on its own, even without ICL, yields impressive results, showcasing its capacity to adhere to specific instructions and complete tasks in a zero-shot scenario. However, it is important to note that integrating these components also increases the associated costs when invoking the LLM.
\begin{table}[h]
  \caption{Ablation studies of LogParser-LLM on Loghub-2k.}
  \label{tab:ablation}
  \footnotesize
  \begin{tabular}{lllllc}
  
    \toprule
    &GA&PA & FGA&FTA&GGD\\
    \midrule
 Drain& 87.2& 40.0 &75.1 &34.4&9.00\\
 Brain& \textbf{96.6}& 40.4 &90.8 &42.7&3.35\\
 LogPPT& 92.3& \textbf{86.5} &89.2&69.5&6.25\\
    \midrule
    GPT-3.5& 91.5 &68.4&86.0   &64.7&5.88\\
    GPT-3.5+ICL+VA& 89.8& 67.2 &86.1 &64.8&5.81\\
 GPT-3.5+ICL+VA+Merge& 90.1& 61.7 &86.9 &59.7&5.50\\
    GPT-4& 92.5 & 75.6 &91.6 &\textbf{75.7}&3.69\\
    GPT-4+ICL+VA+Merge& 91.8& 78.5 &\textbf{92.2}&67.2&\textbf{2.88}\\
  \bottomrule
\end{tabular}
\vspace{-3mm}
\end{table}

\subsection{Evaluation on LogPub}
\footnotetext{**The reason Drain has a GGD of 6618 but a PGD of 23 is that its preprocessing converts all numbers to the variable "<*>". Complicated instance IDs such as "3edec1e4-9678-4a3a" are preprocessed to"<*>edec<*>e<*>-<*>-<*>a<*>a". This results in a significant number of redundant log clusters, leading to a high GGD. However, when calculating the PGD, this is considered a single variable token which is correctly parsed and thus does not contribute to the PGD.}
\textbf{Accuracy and Generalizability}
Results from the expansive logPub dataset are shown in Table~\ref{tab:logpub}. We use LogParser-LLM-C to denote calibrated variants of our method. It is clear that our model, \our, even without granularity calibration, significantly surpasses all baseline methods in GA, FGA, and PTA, marking improvements of 7.8\%, 48.3\%, and 32.0\% compared to the best baseline results. However, PA performance lags, mostly due to granularity nuances complicating the LLM's ability to generate templates that perfectly match annotated labels. A standout point is the consistent performance of our method across the 14 datasets, achieved without domain-specific tweaks, maintaining uniform settings throughout. Upon introducing domain-specific granularity calibration with ICL in LogParser-LLM-C, there is a noticeable boost, especially in template parsing metrics such as PA and FTA. This highlights the reduced discrepancy in the applicability of log parsing achieved through ICL.

\textbf{Granularity Discrepancy Evaluation}
Both Grouping Granularity Distance (GGD) and Parsing Granularity Distance (PGD) are calculated and shown in Table~\ref{tab:logpub}
. PGD is computed using spaces as delimiters for tokenization, representing a lower bound since precise tokenization isn't feasible for such large datasets. This approximation remains valuable for consistent cross-method comparisons.

Unlike message-level GA and PA metrics, which depend on log volume, the proposed metrics avoid template imbalance and provide a clearer performance indicator. For example, in the Linux dataset, Drain’s GA is 68.6 compared to our 53.4. However, Drain's GGD is 30 versus our 10, indicating significantly more effort needed to align Drain's results with the ground truth.

Compared to template-level metrics, GGD and PGD show that smaller GD correlates with higher FGA and FTA. However, FGA and FTA can overly penalize repetitive differences. For example, if a ground truth template "instance: <*>" has many instance IDs not correctly identified as variables, it increases the number of identified templates, skewing precision calculations. GGD and PGD count such differences only once, offering a fairer measurement. For instance, GA and PA for Uniparser on OpenSSH are 0.9 and 0.5, respectively—values that indicate a significant gap compared to other methods and are not informative. Conversely, GGD and PGD for Uniparser on OpenSSH are 15 and 26, respectively, providing an informative and intuitive comparison. This robustness is also observed in HealthApp, OpenStack, and Thunderbird datasets.

\textbf{Evaluation with Different LLMs} By design, our framework is versatile enough to be compatible with any language model that can process individual log messages and accordingly generate log templates. This evaluation's primary objective is to assess the impact of different LLMs on the efficacy and efficiency of our approach.

Our results, as presented in Table~\ref{tab:llms}, demonstrate that using GPT-4 as the LLM template extractor paired with ICL yields optimal performance. However, this comes at the cost of increased computational time due to GPT-4's extensive parameter count. Notably, both granularity calibration and ICL enhance performance and concurrently decrease the number of required LLM calls. This is congruent with our framework's foundational assumption that LLMs can generate nearly perfect log templates. For the fine-tuned LLM, we commenced with the widely-recognized open-source LLM, Llama-2-13b. Despite its performance not being optimal, it remains competitive, closely paralleling the results of prior state-of-the-art semantic-based models like Uniparser and LogPPT. This subpar performance may be attributed to our not having meticulously curated the fine-tuning dataset and its limited size. While a more thoughtfully curated, expansive training dataset and hyperparameter tuning could enhance its performance, this conflicts with our intent: to construct a robust log parser that necessitates minimal human intervention and domain-specific knowledge.

For runtime efficiency, we delineate runtime into two facets: the time required for LLM calls and the time for our log parser's other operations. This distinction is crucial, given that OpenAI service calls depend on service availability and are subject to rate limitations, which complicates consistent performance evaluation. To estimate cumulative processing time, we multiply the average response time (in optimal scenarios) by the total number of LLM calls. Our base algorithm's average runtimes stand at 461.67s and 433.22s. If we disregard potential rate caps, the mean response times are 0.52s for GPT-3.5 and 4.18s for GPT-4. With an average of 272.5 LLM calls, this equates to overall runtimes of 612.31s for GPT-3.5 and 1572.27s for GPT-4. For comparison, Drain, one of the fastest existing methods, averages 483.2s to process about 3.6 million logs. In contrast, a conventional line-by-line LLM parsing approach would necessitate 3.6 million LLM calls, leading to an impractical runtime of approximately 22 days for GPT-3.5, not accounting for other operational overheads. This analysis not only emphasizes our method's competitive efficiency but also its practicality, overcoming the inherent impracticality of existing LLM-based parsers by significantly reducing the number of necessary LLM calls.


\section{Conclusion}
In this study, we introduce LogParser-LLM, a novel approach to log parsing that seamlessly integrates the strengths of Large Language Models (LLMs). Centralizing around a prefix tree and an LLM-based template extractor, LogParser-LLM not only streamlines the extraction of semantically rich log templates but also ensures efficiency through strategic LLM call reductions. While demonstrating compelling results, we also uncovered nuances in parsing granularity, prompting the creation of the Granularity Distance metric. Our rigorous tests on benchmark datasets reveal that LogParser-LLM significantly outshines existing parsers in accuracy and efficiency, demonstrating its potential as a valuable tool for both researchers and practitioners in the field of log analysis.

\begin{acks}
This work was supported by Alibaba Group through Alibaba Research Intern Program.
\end{acks}

\bibliographystyle{ACM-Reference-Format}
\bibliography{main.bib}

\appendix


\section{Additional Discussion}
\subsection{Challenges of Log Parsing in Practice}
The challenges associated with log parsing encompass several key aspects. 

\textbf{Huge Volume}. Modern systems generate vast amounts of log data, which are difficult to manage, store, and analyze. For instance, services like Amazon, Alibaba, and Facebook generate billions of visits per day, each creating multiple log entries~\cite{MetaReports2023, gangneux2019rethinking}. Log parsing, along with tasks like anomaly detection and root cause analysis, is crucial for minimizing system downtime and financial loss~\cite{fu2009execution, rao2011identifying, gurumdimma2015towards}. The requirement for real-time, streaming log parsing makes handling such vast volumes challenging.

\textbf{Constantly Evolving}. Systems and technologies continuously evolve, leading to changes in log entry types, formats, structures, and content. New features and components introduce novel log formats, necessitating updates to log templates for accurate parsing. Without timely template updates, parsing algorithms may fail to extract relevant information, leading to inaccuracies and incomplete analysis. Proactively updating log templates ensures effective parsing and adaptation to dynamic log generation.

\textbf{Diverse Sources}. Logs from different systems often have diverse formats, posing a challenge for log parsing algorithms. Each system's unique log format can vary significantly in structure, syntax, and content. Effective log parsing algorithms must generalize to handle various formats without relying on system-specific rules or assumptions.

\subsection{Insights and Opportunities of Log Parsing with LLMs}

The relentless growth in log volumes, the ever-evolving nature of logs, and the vast diversity in log sources have presented daunting challenges in the realm of log parsing. Syntax-based parsers, while efficient, often grapple with the dynamic nuances introduced by log evolution and diverse sources. LLMs, with their deep semantic understanding and adaptability, are poised as a promising solution but need prompt tuning and optimization to handle vast volumes.

Moreover, the vital role of log data in modern systems underscores the need for log parsing tools that embody certain foundational principles. In practice, a log parser must be \textbf{Accurate}, ensuring accurate interpretation of every piece of information. \textbf{Efficiency} is paramount to handling the voluminous log data churned out by popular platforms and sprawling systems. The parser's \textbf{Evolvability} will be its asset, granting it the flexibility to keep pace with system updates and new feature integrations. To confront the multifarious log formats from diverse sources, it is imperative that a parser is \textbf{Generalizable}, ensuring it doesn't rely too heavily on system-specific constructs.

Furthermore, for real-time responsiveness, the parser needs to operate in an \textbf{Online} manner. This demands the tool's agility to adapt and recalibrate as new log entries stream in. Addressing the challenge of different granularities in log parsing is also of utmost importance. Ensuring the capability to \textbf{Calibrate Granularity} provides flexibility in parsing logs, given that a single log can be interpreted in multiple, yet reasonable, ways based on granularity.

Given the unique strengths and challenges of each approach, a compelling motivation emerges: to amalgamate the adaptability and depth of LLMs with the efficiency intrinsic to syntax-based parsers. This convergence promises a robust and versatile log parsing solution, aptly suited to address both present and future challenges in log management.

\section{Existing metrics} \label{app_metrics}
\textbf{Grouping Accuracy (GA)}
GA measures the ratio of correctly grouped log messages. A message is considered correctly grouped if and only if its template group is exactly aligned with ground truth grouping.

\textbf{Parsing Accuracy (PA)}
PA assesses the ability to extract templates accurately, critical for tasks like anomaly detection. it is the fraction of messages parsed correctly, meaning all template and variable tokens are identified accurately.

\textbf{F1 score of Grouping Accuracy (FGA)}
FGA is a template-level metric that evaluates the fraction of correctly grouped templates. Using the true number of templates (\(N_g\)), parsed templates (\(N_p\)), and correctly parsed templates (\(N_c\)), we calculate the Precision (\(PGA = \frac{N_c}{N_p}\)) and Recall (\(RGA = \frac{N_c}{N_g}\)) of Grouping Accuracy. FGA is their harmonic mean.

\textbf{F1 score of Template Accuracy (FTA)} FTA is the harmonic mean of \textbf{Recall of Template Accuracy (RTA)} and 
\textbf{Precision of Template Accuracy (PTA)}. Like FGA, FTA evaluates correct template identification at the template level. A template is correct if log messages with the same parsed template share the same ground-truth template and the parsed template matches the ground-truth template exactly. Using $\hat{N_c}$ to denote the number of templates identified accurately by a parser, PTA is then given by $\frac{\hat{N_c}}{N_p}$, and RTA by $\frac{\hat{N_c}}{N_g}$, allowing us to compute FTA as $2 \times \frac{PTA \times RTA}{PTA + RTA}$.

\section{Additional Implementation Details}
\textbf{Fine-tuning Settings} The llmama-2-13b model was finetuned on a server equipped with 8 Tesla A100 80GB GPUs using the Hugging Face Transformers package. The model was finetuned for 50 epochs with 32 samples for each dataset. During inference, we utilized DeepSpeed\cite{rasley2020deepspeed} with 8-bit quantization to expedite the inference process on a single Tesla A100 80GB GPU. Additionally, the model was fine-tuned using LoRA~\cite{hu2021lora} with rank $r$ set to 64. For optimization, we employed the AdamW optimizer with an initial learning rate of 2e-4, which was linearly scheduled down to 0. The batch size was 16.

\textbf{Prompts} We demonstrate the final prompt used for the ICL-based method in Figure~\ref{fig:promp2} and the fine-tuning-based method in Figure~\ref{fig:prompt3}. The prompts for automatic template merge check and verification are shown in Figure~\ref{fig:prompt4} and Figure~\ref{fig:prompt5}, respectively.

%
%
%
%
%
%
%
\begin{figure}[ht]
\centering
\fbox{
  \parbox{\linewidth}{
    \ttfamily
    """As a log parser, your task is to analyze logs and identify dynamic variables. These variables are distinct from static parts, which are hardcoded sections in the logging code. The categories of dynamic variables are concluded as:
\\

Object ID (OID): Includes variables like session IDs and user IDs.

Location Indicator (LOI): Path information, URIs, and IP addresses.

Object Name (OBN): Domain names, task names, job names.

Type Indicator (TID): Category for type indicators.

Switch Indicator (SID): Category for switch indicators (only numerical ones).

Time/Duration of an Action (TDA): Timespan or duration of actions.

Computing Resources (CRS): Memory, disk space, number of bytes.

Object Amount (OBA): Number of errors, nodes, etc.

Status Code (STC): Error codes (only numerical ones).

Other Parameters (OTP): All other types of variables.
\\

To parse the logs, substitute dynamic variables with their respective category tokens, denoted by <XXX>. Everything outside the <XXX> should remain exactly unchanged! Do not fix any typo! If a variable comprises several smaller, fine-grained variables, don't dissect it. Instead, replace the entire compound variable with a single <XXX> token. Do not substitute all content in the log as a variable; only genuine dynamic variables should be replaced.
\\

Examples:

Log: \{example log message\}

Parsed Log: \{example template\}

...

Log: \{log to be parsed\}

Parsed Log: """ 
  }
}
\caption{Variable-aware Prompt for Log Parsing}
\label{fig:promp2}
\end{figure}

\begin{figure}[ht]
\centering
\fbox{
  \parbox{\linewidth}{
    \ttfamily
        """Below is an instruction that describes a task. Write a response that appropriately completes the request

        \#\#\# Instruction:
        
        Analyze the input log and identify dynamic variables. Substitute dynamic variables with <XXX>. 
        
        \#\#\# Input:
        
        \{log\}
        
        \#\#\# Response:
        
        \{template\}

        \#\#\# End """

  }
}
\caption{Prompt for fine-tuning}
\label{fig:prompt3}
\end{figure}
\begin{figure}[ht]
\centering
\fbox{
  \parbox{\linewidth}{
    \ttfamily
    """
    
 \{task prompt in Figure~\ref{fig:promp2}\}
 \\
 
 Given the following logs, output the parse result for each of them first, then determine whether they are instances from the same event template. The output should use the following format:
\\ 

EventTemplate\_1: \{parse result for Log\_1\}

EventTemplate\_2: \{parse result for Log\_2\}

...

EventTemplate\_N: \{parse result for Log\_N\}
\\

Reason: \{brief reason whether they should be unified\}
\\

Answer: \{"Yes" or "No"\}
\\

Unified Template: \{one unified template if yes. Make sure there are static parts in the template. "None" if the anwser is no\}

"""

  }
}
\caption{Prompt for Merge Verification}
\label{fig:prompt4}
\end{figure}
\begin{figure}[ht]
\centering
\fbox{
  \parbox{\linewidth}{
    \ttfamily
    """
 \{task prompt in Figure~\ref{fig:promp2}\}
\\ 
Does the template: "\{merged\_template\}" apply to the following logs? Please answer with yes or no.
\\    

    Answer:

"""

  }
}
\caption{Prompt for Merge Checking}
\label{fig:prompt5}
\end{figure}
\end{document}